\documentclass[aps,pr,floatfix,twocolumn]{revtex4}
\usepackage{graphicx,amsmath,units,xspace}

\graphicspath{{figures/}}   

\newcommand{\micron}{\ensuremath{\unit{\mu m}}\xspace}

\renewcommand{\vec}[1]{{\mathbf #1}}
\newcommand{\avg}[1]{\left< #1 \right>}
\newcommand{\abs}[1]{\vert #1 \vert}
\newcommand{\order}[1]{{\cal O}\left( #1 \right)}

\newcommand{\vecr}{\ensuremath{\vec{r}}\xspace}

\newcommand{\vecG}{\vec{\Gamma}}
\newcommand{\ham}{{\mathcal H}}
\newcommand{\hamiltonian}{\ensuremath{\ham(\vecG)}\xspace}
\newcommand{\pot}{V}
\newcommand{\potential}{\ensuremath{\pot(\{q_i\})}\xspace}
\newcommand{\kinetic}{\ensuremath{K(\{p_i\})}\xspace}
\newcommand{\vecB}{\vec{B}(\vecG)}
\newcommand{\force}{\vec{F}}
\newcommand{\Th}[1]{T_h^{(#1)}}

\begin{document}                                             
                                                             
\title{Configurational Temperatures and Interactions in Charge-Stabilized
  Colloid}

\author{Yilong Han}

\affiliation{Department of Physics and Astronomy, University of Pennsylvania,
  209 South 33rd St., Philadelphia, PA 19104}                

\author{David G. Grier}

\affiliation{Department of Physics and Center for Soft Matter Physics, 
  New York University, 4 Washington Place, New York, NY 10003}

\date{\today}                                               
                                                              
\begin{abstract}
  A system's temperature 
  can be defined in terms of its constituents'
  instantaneous positions rather than their momenta. 
  Such configurational temperature definitions offer substantial
  benefits for
  experimental studies of soft condensed matter systems, most notably
  their applicability to overdamped systems whose instantaneous
  momenta may not be accessible.
  We demonstrate that the configurational temperature formalism
  can be derived from the classical hypervirial theorem,
  and introduce a hierarchy of hyperconfigurational temperature
  definitions, which are particularly well suited for experimental studies.
  We then use these analytical tools
  to probe the electrostatic
  interactions in monolayers of charge-stabilized colloidal spheres
  confined by parallel glass surfaces.
  The configurational and hyperconfigurational temperatures, together
  with a novel thermodynamic sum rule, provide previously lacking
  self-consistency tests for interaction measurements based on digital
  video microscopy, and thereby cast new light on controversial
  reports of confinement-induced like-charge attractions.
  We further introduce a new method for measuring the pair potential 
  directly that
  uses consistency of the configurational and hyperconfigurational 
  temperatures as a set of constraints for a model-free search.
\end{abstract}

\maketitle

\section{Generalized Temperature Definitions}
\label{sec:theory}

A variety of thermodynamic temperature definitions complementary 
to the classic kinetic definition have been derived recently \cite{Rugh97,Butler98,Jepps00}.
The most general form, proved in Ref.~\cite{Jepps00} and \cite{Rickayzen01}, is: 
\begin{equation}  
  \label{eq:Tgeneral}
  k_B T = 
  \frac{\avg{\nabla \hamiltonian \cdot \vecB}}{\avg{\nabla \cdot \vecB}},
\end{equation}
where
$\vecG = \{\Gamma_{1}, \cdots , \Gamma_{6N}\} = 
\{q_1, \cdots , q_{3N}, p_1, \cdots , p_{3N}\}$ 
is the set of $3N$ generalized
coordinates $q_i$ and $3N$ conjugate momenta $p_i$ describing the 
$6N$-dimensional phase space of an $N$-particle system in equilibrium,
and angle brackets $\avg{\cdots}$ represent an
ensemble average.
The system's Hamiltonian, 
$\hamiltonian = \kinetic + \potential$,
consists of the potential energy $\kinetic = \sum_{i=1}^{3N} p_i^2/(2m)$ and
the conservative $N$-body potential \potential.
The most noteworthy feature of Eq.~(\ref{eq:Tgeneral}) is that
$\vecB$ can be \emph{any}
continuous and differentiable vector in phase space \cite{Rickayzen01}.
Choosing $\vecB = \{0, \cdots, 0, p_1, \cdots, p_{3N}\}$
yields the familiar equipartition theorem,
\begin{equation}
  \label{eq:equipartition}
  k_B T = \avg{\frac{1}{N} \, \sum_{j=1}^N \frac{p_j^2}{m_j}},
\end{equation}
where $m_j$ is the mass of the $j$-th particle.
Choosing instead
$\vecB = -\nabla \potential$ yields
\begin{equation}
  \label{eq:Tconfig}
  k_B T_{config} = \frac{\avg{\nabla \pot \cdot \nabla \pot}}{
    \avg{\nabla^2 \pot}},
\end{equation}
which depends only on the objects' positions and not their momenta.
This has come to be called the configurational temperature.

The configurational temperature's independence of the particles' momenta has 
important ramifications for experimental studies of systems such as colloidal 
suspensions whose configurations are easily measured but whose momenta are not.
The experiments described in this 
Article exploit properties of the configurational temperature
to obtain new insights into the interactions between charge stabilized
colloidal particles.  
When combined with thermodynamic sum rules, this formalism provides
previously lacking
thermodynamic self-consistency tests for measurements of the particles'
effective pair potentials.
The same formalism also can be used to measure
pair potentials in soft-matter systems directly, thereby
bypassing questions of interpretation raised in previous
studies, and yielding comparably accurate results with substantially
less data.

Section~\ref{sec:tconfig} provides an overview of several consequences of
Eq.~(\ref{eq:Tgeneral}).
These are extended in Sec.~\ref{sec:hyperconfigurational} to a
hierarchy of hyperconfigurational temperatures that lend themselves
particularly nicely to experimental studies.
Section~\ref{sec:hypervirial} provides an alternate foundation for
the entire configurational temperature formalism in the classical
hypervirial theorem.
Consistency among the myriad temperature definitions is possible only
if the assumptions underlying their derivations all are satisfied
simultaneously.
Applying these definitions to experimental data, as described
in Sec.~\ref{sec:experiment}, therefore probes the nature of the
system's inter-particle interactions.
These data were obtained from digital video microscopy
measurements on  monolayers of charge-stabilized
colloidal spheres dispersed in water between parallel glass
surfaces.  
Not only does this system lend itself naturally to computing 
the configurational temperature, but the results also help 
to resolve a long-standing controversy
regarding the nature of charged colloids' interactions in confined
geometries.
These results are summarized in Sec.~\ref{conclusion}.

\section{Configurational Temperatures}
\label{sec:tconfig}
The most general configurational temperature definition,
Eq.~(\ref{eq:Tconfig}), requires knowledge of the full
$N$-body potential, which typically is not known for
experimental systems.
A more experimentally accessible form emerges
if the particles interact via pairwise additive potentials,
$u_{ij}(\vec{r}_i-\vec{r}_j)$.
In this case, we can interpret gradients of the $N$-body potential,
\begin{align}
  \nabla \potential & = 
  \left(
    \frac{\partial}{\partial q_1},\cdots,\frac{\partial}{\partial q_{3N}} 
  \right) \,
  \sum_{i=1}^{N-1} \sum_{j=i+1}^N u_{ij}(\vecr_i-\vecr_j) \nonumber \\
  & = - ( \force_1, \dots, \force_N )
  \label{eq:pairwise}
\end{align}
as components of the net force acting on each of the particles due to their
interactions with their neighbors, where
\begin{equation}
  \label{eq:netforce}
  \force_j = \nabla_j \sum_{i \neq j} u_{ij}(\vecr_i - \vecr_j)
\end{equation}
is the force acting on the $j$-th particle.
The temperature then may be written as \cite{Jepps00}
\begin{equation}
  \label{eq:TconfigF}
  k_B T_{conF} =\frac{\avg{\sum_{j=1}^N \force_j^2}}{
    \avg{-\sum_{j=1}^N \nabla_j \cdot \force_j}}.
\end{equation}
It makes sense that the temperature should be reflected in the instantaneous
distribution of forces
because objects explore more of their potential energy landscape as the temperature increases.

All definitions of the temperature derived
from Eq.~(\ref{eq:Tgeneral})
involve approximations of $\order{1/N}$, and so only are
valid in the thermodynamic limit, $N \to \infty$.
Dropping additional terms of $\order{1/N}$ in the derivation of
Eq.~(\ref{eq:TconfigF}) can be justified for systems with short-ranged
interactions and leads to another equivalent temperature definition \cite{Jepps00}:
\begin{equation}
  \label{eq:Tconfig1}
  k_B T_{con1} = \avg{
    \frac{ \sum_{j=1}^N \force_j^2}{-\sum_{j=1}^N \nabla_j \cdot \force_j}}.
\end{equation}
Neglecting further higher-order terms yields still another form \cite{Han04a},
\begin{equation}
  \label{eq:Tconfig2}
  k_B T_{con2} = \avg{\frac{-\sum_{j=1}^N \nabla_j \cdot \force_j}{
      \sum_{j=1}^N \force_j^2}}^{-1}.
\end{equation}

\section{Hyperconfigurational temperatures}
\label{sec:hyperconfigurational}
Choosing $\vecB = \{q_1^s, \cdots, q_{3N}^s, 0, \cdots, 0\}$ with 
$s = 1, 2, 3 \cdots$ in Eq.~(\ref{eq:Tgeneral}) yields a 
hierarchy of so-called hypervirial temperatures \cite{Hirschfelder60}, 
which reduce to the Clausius's virial temperature for $s = 1$. 
By the same token, we propose that
$\vecB = \{F_1^s, \cdots, F_{3N}^s, 0, \cdots, 0\}$ with $s > 0$, 
yields the set of ``hyperconfigurational temperatures'' \cite{Han04a},
\begin{equation}
  \label{eq:hyperT}
  k_B \Th{s} = \frac{\avg{\sum_{i=1}^{3N} F_i^{s+1}}}{
    \avg{-s\sum_{i=1}^{3N}F_i^{s-1} \partial_i F_i}},
\end{equation}
of which $\Th{1}$ is equivalent to the standard configurational temperature. 
Here, $F_i$ is the magnitude of the $i$-th element in the set of
$3N$ components of the forces on the $N$ particles.  
Because $F_i$ is non-negative,
$\Th{s}$ is well defined for any positive real value of $s$.
Negative values of $s$ would yield diverging temperatures because
at least some of the $F_i$ will be vanishingly small for any system
substantially larger than the range of interactions.

A simple example motivates introducing this new hierarchy of definitions. 
If, for example, a system is characterized by Coulomb pair interactions,
$u(r) = 1/r$ in $d = 3$ dimensions, each term of the denominator,
$\nabla_r^2 u(r) = r^{1-d} \, \partial_r(r^{d-1}\partial_r u(r))$, 
of Eqs.~(\ref{eq:TconfigF}) and (\ref{eq:Tconfig1}) vanishes.
Consequently, the associated configurational temperature 
definitions in
Eqs.~(\ref{eq:TconfigF}),~(\ref{eq:Tconfig1}) and~(\ref{eq:Tconfig2}) diverge
unphysically. 
The hyperconfigurational temperatures, by contrast, are still well defined
with
$\partial_x F_x = (1-3x^2/r^2)/r^3$, and  
$F_x^{s-1} \partial_x F_x + F_y^{s-1} \partial_y F_y + F_z^{s-1}\partial_z F_z\ne 0$ 
for $s \ne 1$.
Consequently, the hyperconfigurational temperatures should apply
to any system whose pair potential is continuous and differentiable.
This suggests that they will be useful for studying systems whose interactions
are not known \emph{a priori}.

Additional useful results emerge for systems such as colloidal monolayers
whose interactions are isotropic.
In this case, the Cartesian coordinates may be analyzed independently
\begin{eqnarray}
  k_B T_{conF} & = & \frac{\avg{\sum_{j=1}^N \force_j^2}}{
    \avg{-\sum_{j=1}^N \nabla_j \cdot \force_j}}
   = \frac{\avg{\sum_{j=1}^N (F_{j,x}^2 + F_{j,y}^2)}}{
    \avg{- \sum_{j=1}^N (F^\prime_{j,x} + F^\prime_{j,y})}} \nonumber\\
  & = &\frac{\avg{\sum_{j=1}^N F_{j,x}^2}}{
    \avg{- \sum_{j=1}^N F^\prime_{j,x}}}
  = \frac{\avg{\sum_{j=1}^N F_{j,y}^2}}{\avg{- \sum_{j=1}^N F^\prime_{j,y}}}.
  \label{eq:tcomponents}
\end{eqnarray}
Setting $\vecB = \partial_x \potential$ or 
$\partial_y \potential$ in Eq.~(\ref{eq:Tgeneral}) leads to same results.
We will refer to the two terms in Eq.~(\ref{eq:tcomponents}), as well as
analogous results for other temperature definitions, as the
Cartesian components of the configurational temperature, $T_x$ and
$T_y$, respectively.

\section{Derivation from the hypervirial theorem}
\label{sec:hypervirial}
A remarkable consequence of Eq.~(\ref{eq:Tgeneral})
is that \emph{any} vector field
$\vecB$ that depends only on the $3N$ configurational degrees of
freedom gives rise to a functionally distinct but thermodynamically
equivalent definition of the temperature that depends
only on configurational coordinates.
Here we show that this insight emerges transparently from
the hypervirial theorem \cite{Hirschfelder60,Marc85}, and that
Eq.~(\ref{eq:Tgeneral}) itself can be derived from 
this starting point.

The Hamiltonian equation of motion for an arbitrary
dynamical variable, $f(\vecG,t)$, is
\begin{equation}
  \label{eq:hamiltonian}
  \frac{df}{dt} = \frac{\partial f}{\partial t} + \{f,\hamiltonian\},
\end{equation}
where 
\begin{equation}
  \label{eq:poissonbracket}
  \{f,\hamiltonian\} = \sum_{i=1}^{3N} \, \left[
      \frac{\partial \hamiltonian}{\partial p_i} \,
      \frac{\partial f}{\partial q_i} -
      \frac{\partial \hamiltonian}{\partial q_i} \,
      \frac{\partial f}{\partial p_i}\right]
\end{equation}
is the Poisson bracket of $f$ and $\hamiltonian$.
Provided that (1) $f$ does not depend explicitly on time and (2)
$f(\vecG)$ remains bounded in the course of time,
the time average of Eq.~(\ref{eq:hamiltonian}) yields the
classical hypervirial theorem,
\begin{equation}
  \label{eq:HVT}
  \avg{\{f,\hamiltonian\}} = 0
\end{equation}
or, equivalently,
\begin{equation}
  \label{eq:HVT2}
  \sum_{i=1}^{3N} \avg{\frac{\partial \mathcal{H}}{\partial p_i} 
  \frac{\partial f}{\partial q_i}}
  = \sum_{i=1}^{3N} \avg{\frac{\partial \hamiltonian}{\partial q_i} \,
    \frac{\partial f}{\partial p_i}}.
\end{equation}

If we restrict $f$ to be a homogeneous first-degree
function of the momenta, 
\begin{equation}
  \label{eq:HVT3}
  f(\vecG) = p_j \, Q(\{q_i\}),
\end{equation}
while still allowing $Q$ to be an arbitrary function of coordinates, 
the right hand side of Eq.~(\ref{eq:HVT2}) becomes
\begin{equation}
  \label{eq:HVT4}
 \sum_{i=1}^{3N} \avg{\frac{\partial \hamiltonian}{\partial q_i} \,
    \frac{\partial f}{\partial p_i}}
  = \avg{Q \frac{\partial \hamiltonian}{\partial q_j}}
  = \avg{Q \, \frac{\partial \potential}{\partial q_j}}.
\end{equation}
The coordinates and momenta are statistically
uncorrelated in equilibrium, so that
\begin{align}
  \sum_{i=1}^{3N} \avg{\frac{\partial \hamiltonian}{\partial p_i} \,
    \frac{\partial f}{\partial q_i}}
  & = \sum_{i=1}^{3N} \avg{\frac{\partial \hamiltonian}{\partial p_i} \, p_j}
  \avg{\frac{\partial Q}{\partial q_i}} \\
  & = \avg{p_i \, \frac{\partial \hamiltonian}{\partial p_i}}
  \avg{\frac{\partial Q}{\partial q_i}},
  \label{eq:HVT5}
\end{align}
where terms with $i \ne j$ vanish.
Now, $\avg{p_i\frac{\partial \ham}{\partial p_i}} = k_B T$ 
is the standard kinematic definition of the temperature.
Substituting this, Eq.~(\ref{eq:HVT4}) and Eq.~(\ref{eq:HVT5}) into
Eq.~(\ref{eq:HVT2}) yields
\begin{equation}
  \label{eq:HVT6}
  k_B T = \avg{Q \, \frac{\partial \potential}{\partial q_j}}
  \avg{\frac{\partial Q}{\partial q_i}}^{-1},
\end{equation}
which is a special case of Eq.~(\ref{eq:Tgeneral}) 
when $\vecB = \mathbf{B}(\{q_i\})$.

Hirschfelder~\cite{Hirschfelder60} chose $Q = q_i^s$
to obtain a hierarchy of hypervirial temperatures
characterized by the index $s$.
Choosing instead
$Q = F_i^s$ yields the hyperconfigurational 
temperatures in Eq.~(\ref{eq:hyperT}).

A similar line of reasoning provides a straightforward
derivation of the most general result, Eq.~(\ref{eq:Tgeneral}).
We begin by choosing $Q = q_i$ in Eq.~(\ref{eq:HVT6}) to obtain
$k_B T = \avg{q_i  \frac{\partial \ham}{\partial q_i}}$,
which is equivalent to the Clausius virial theorem,
$k_B T = \avg{r_i F_i}$, for systems with pairwise additive interactions.
Next, we choose $f(\vecG) = q_i \, P(\{p_i\})$, where $P$ is an arbitrary
function of the momenta, and substitute into Eq.~(\ref{eq:HVT2}).
The following steps are analogous to those used in deriving Eq.~(\ref{eq:HVT6})
and yield
\begin{equation}
  \label{eq:HVT7}
   k_B T = \avg{P \, \frac{\partial \kinetic}{\partial p_j}}
  \avg{\frac{\partial P}{\partial p_i}}^{-1}.
\end{equation}
Combining this with Eq.~(\ref{eq:HVT6}) yields
\begin{align}
  k_B T & = \frac{\avg{QP \left(
      \frac{\partial \potential}{\partial q_i} + \frac{\partial \kinetic}{\partial p_i}
      \right)}}{\avg{
      P \, \frac{\partial Q}{\partial q_i} + Q \, \frac{\partial P}{\partial p_i}}} \\
  & = \frac{\avg{QP \, \nabla_i \hamiltonian}}{\nabla_i (QP)}, \label{eq:QED}
\end{align}
where, once again, we have exploited the statistical independence of the coordinates
and momenta.
Because this holds for any choice of $Q$ and $P$, it holds for any sum of products of the
form $\sum_m Q_m P_m$.
Thus, Eq.~(\ref{eq:QED}) is equivalent to Eq.~(\ref{eq:Tgeneral}) for any
choice of $\vecB$ whose components can be expressed as Taylor series in the
coordinates and momenta.

\section{Application to Colloidal Dispersions}
\label{sec:experiment}

Because the positions of microscopic particles, such as atoms, are hard to measure, 
experimental applications of the configurational temperature have been 
relatively limited~\cite{Han04a}. 
Colloidal dispersions differ from many experimental systems in that their 
constituents' motions can be tracked quite easily. 
Unlike most macroscopic model systems, they can be thoroughly equilibrated 
through the particles' intimate contact with the suspending fluid.
Furthermore, an effective potential between two particles often can be defined. 
For this reason, colloidal dispersions offer an almost ideal experimental test-bed for
studying the configurational temperature and related ideas.
These ideas, in turn, provide valuable new tools for assessing the
microscopic state of this important state of matter.

This section describes methods for estimating 
the configurational and hyperconfigurational
temperatures of colloidal dispersions from digital video microscopy
measurements of their microscopic dynamics.
Emphasis is placed on how to account for inevitable experimental
errors in systems of limited size whose interactions may not be
fully characterized \emph{a priori}.
In particular, we analyze the structure and dynamics of dilute 
monolayers of charge-stablized
colloidal spheres confined by planar surfaces, on the one hand using convergence of
different temperature definitions as a thermodynamic self-consistency test for
independently measured effective pair potentials, 
and on the other as a new approach to measuring the pair potential directly.

\subsection{Interactions in charge-stabilized colloid}
Considerable attention has been focused in recent years on colloidal interactions,
particularly in light of experimental observations that 
challenge theoretical predictions.
For example, charged colloidal spheres of diameter $\sigma$ and charge
number $Z$ dispersed in an electrolyte interact with
each other directly through their Coulomb repulsion and also indirectly through
their influence on the surrounding distribution of simple, atomic-scale ions.
Poisson-Boltzmann mean-field theory predicts an overall 
screened-Coulomb repulsion \cite{Russel89} of the form
\begin{equation}
  \label{eq:DLVO}
  \beta u(r) = Z^2 \lambda_B \, 
  \frac{\exp(\kappa \sigma)}{\left(1 + \frac{\kappa \sigma}{2}\right)^2} \, 
  \frac{\exp(-\kappa r)}{r},
\end{equation}
where $\beta^{-1} = k_BT$ is the thermal energy scale at absolute temperature $T$, 
$r$ is the spheres' center-to-center separation, 
and $\lambda_B = e^2/(4\pi \epsilon k_B T)$ is the Bjerrum length 
in a medium of dielectric constant $\epsilon$.  
If the electrolyte has a \textit{total} concentration $c$ of monovalent ions, 
then the Debye-H\"uckel screening length,
$\kappa^{-1} = 1/\sqrt{4 \pi \lambda_B c}$,
sets the range of the effective electrostatic interaction in the mean-field
approximation.
This is an effective interaction because it results from an
average over the simple ions' degrees of freedom.
When viewed in this light, it is not surprising that measurements might
differ from predictions based on Eq.~(\ref{eq:DLVO}).
More surprising is that like-charged colloidal spheres appear to 
\emph{attract}
each other under some circumstances, in qualitative disagreement with
Eq.~(\ref{eq:DLVO}).
One of the goals of the present study is to apply the configurational temperature
formalism to resolve some of the outstanding questions regarding these
anomalous like-charge attractions.

Unlike atoms, which travel ballistically within the potential energy landscape
established by inter-atomic interactions, colloidal particles are immersed in a
viscous fluid that randomizes their trajectories over intervals longer than
$\tau = \beta m D$, the momentum relaxation time.
Given a typical colloidal diffusion coefficient $D \approx 1~\unit{\micron^2/sec}$,
$\tau \approx 1~\unit{\mu sec}$.
Consequently colloidal suspensions' microscopic temperatures are not easily 
monitored with the usual kinetic definition of the temperature, 
Eq.~(\ref{eq:equipartition}).
The configurational temperature, by contrast, can be measured from snapshots
and so provides an ideal alternative.

The fluid also acts as an intimately coupled heat bath whose heat capacity
vastly exceeds the colloidal particles'.
Consequently, the dispersion's thermodynamic temperature 
is all but guaranteed to be the fluid's,
which is readily monitored with standard techniques.
It is natural, therefore, to compare estimates of the configurational temperature
based on microscopic dynamical measurements with this bulk thermodynamic
temperature.


Our samples consist of negatively charged silica spheres $\sigma = 1.58~\micron$ in
diameter (Duke Scientific Lot. 24169) dispersed in water and 
confined within a slit pore of height $H$
formed between a glass microscope slide and a cover slip.
The glass surfaces also develop
large negative charge densities in contact with water \cite{Behrens01b},
which repel the spheres and prevent them from sticking under the influence 
of van der Waals attraction. 
Silica spheres are roughly twice as dense as water and
sediment onto the lower wall in a matter of seconds.
The low-concentration samples used in this study thus
form a dilute monolayer once they reach equilibrium.
Reservoirs of
mixed-bed ion exchange resin help to maintain a total ionic strength 
around $c = 5 \times 10^{-6}~\unit{M}$ in the 
$1 \times 4~\unit{cm^2}$ visible sample area. 

The hermetically sealed sample is allowed to equilibrate at ambient 
temperatures on the stage of a microscope. 
The particles' motions are imaged with a CCD (charge-coupled device) camera 
and video taped at 30 frames/sec before being digitized.
Standard methods of digital video analysis \cite{crocker96} identify the
particles in each video frame and report their locations in the plane with
a resolution of 30~\unit{nm}. 
The resulting distribution,
\begin{equation} 
  \label{eq:trajectory}
  \rho(\vecr, t) = \sum_{j=1}^{N(t)} \delta(\vecr - \vecr_j(t)),
\end{equation}
of $N(t)$ particles in the field of view at time $t$
provides detailed information regarding the particles' 
dynamics under the combined influences of random thermal forces and their 
mutual interactions. 
Distilling this information into an easily interpreted form
requires further analysis.

One of the most commonly used tools for analyzing
colloidal microscopy data is
the radial distribution function, $g(r)$,
which is computed as
\begin{equation}
  \label{eq:gr}
  g(r) = \frac{1}{n^2} \, \avg{\rho(\vecr - \vecr', t) \rho(\vecr', t)},
\end{equation}
where $n = \avg{\rho} = N/A$ is the areal density of particles
in a field of view containing $N = \avg{N(t)}$ particles. 
Angle brackets in Eq.~(\ref{eq:gr}) denote averages over both angle and time.
A typical example appears in the inset to Fig.~\ref{fig:u(H=9)}.

\subsection{Colloidal interaction measurements}
For a system with radially symmetric pairwise-additive interactions,
the radial distribution function can be related to the pair potential, $u(r)$.
As a starting point, we introduce the potential of mean force,
\begin{equation}
  \label{eq:meanforce}
  w(r) \equiv - k_B T \ln g(r),
\end{equation}
which reduces to the pair potential in the dilute limit:
$\lim_{n \to 0} w(r) = u(r)$.
Crowding at higher concentrations induces many-body correlations
that appear as oscillations both in $g(r)$ and also in $w(r)$.
Such oscillations arise even in systems with monotonically repulsive
interactions, and could be mistaken for structure in $u(r)$.
Reducing the concentration to avoid many-body correlations often is not
practical given the competing requirements of tracking particles accurately
(which favors high magnifications \cite{Crocker96thesis})
and amassing adequate statistics in $g(r)$
(which favors a wide field of view and thus low magnification 
\cite{Behrens01a}).

No exact closed-form 
relationship between $g(r)$ and $u(r)$ is known for systems 
at finite concentrations.
Fortunately, Henderson's uniqueness theorem \cite{Henderson74}
guarantees that any
trial potential that reproduces $g(r)$ uniquely describes the 
system's interactions,
provided these are indeed pairwise additive.
In light of this, two complementary methods for inverting $g(r)$ to obtain $u(r)$ 
have been introduced.
The inverse Brownian dynamics \cite{kepler94} 
and inverse Monte Carlo \cite{rajagopalan97} 
techniques compare the experimental
radial distribution function with results 
obtained from numerical simulations based
on experimental conditions and trial potentials until a match is obtained.
The liquid structure inversion technique \cite{carbajaltinoco96}
uses the Ornstein-Zernike (OZ) integral equation \cite{mcquarrie00} to 
correct for many-body correlations directly.
This is the approach we will adopt.

The OZ equation
describes the evolution of many-body correlations from a hierarchy 
of pairwise interactions. 
Truncating the hierarchy yields approximate but analytically tractable relationships
between $g(r)$ and $u(r)$.
Among these, the hypernetted chain (HNC) approximation is found to
be accurate for ``soft'' potentials while the Percus-Yevick (PY)
approximation
is more accurate for short-ranged interactions.
For two-dimensional systems,
the pair potential can be evaluated
in these approximations as \cite{chan77}
\begin{equation}
  \label{eq:oz}
  u(r) = w(r) + 
  \begin{cases}
    k_B T \, n \, I(r) & \text{(HNC)}\\
    k_BT \, \ln (1 + n\, I(r)) & \text{(PY)} 
  \end{cases}
\end{equation}
where the convolution integral,
\begin{equation}
  \label{eq:ir}
  I(r) = \int_A \left[ g(r') - 1 - n I(r) \right]
  \left[ g(\abs{{\vecr'- \vecr}}) - 1  \right] d^2r',
\end{equation}
is solved iteratively, starting with $I(r) = 0$.

\begin{figure}[t!]
  \centering
  \includegraphics[width=\columnwidth]{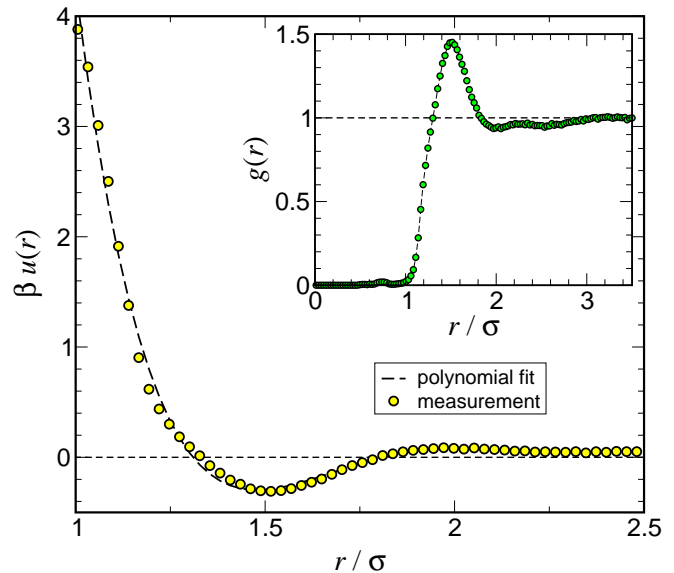} 
  \caption{Inset: Radial distribution function $g(r)$ for a monolayer of 
    $\sigma = 1.58~\micron$
    diameter silica spheres sedimented onto a layer at areal density
    $n \sigma^2 = 0.0684$ above a glass surface in a parallel-plate slit pore
    of height $H = 9~\micron$.  Main plot: Effective pair potential obtained from
    $g(r)$ using Eqs.~(\ref{eq:oz}) and (\ref{eq:ir}).  The smooth dashed
    curve is a fifth-order polynomial fit.}
  \label{fig:u(H=9)}
\end{figure}

Provided the particles' concentration is not too high, and that their
interactions are pairwise additive and neither too
long-ranged nor too short, and that $g(r)$ is known with sufficient accuracy,
both approximations should yield equivalent results for $u(r)$.
Unfortunately, consistency does not guarantee accuracy \cite{Behrens01a}, 
and additional checks are necessary to ensure that pair potentials
obtained with Eqs.~(\ref{eq:oz}) and (\ref{eq:ir}) are meaningful.
This is all the more important if such measurements as the
example in Fig.~\ref{fig:u(H=9)},
run counter to long-established theory \cite{Han03b}.
In this particular case, the $0.3~k_BT$ deep minimum at $r = 1.5~\sigma$
is inconsistent with the purely repulsive interaction described
by Eq.~(\ref{eq:DLVO}).
Such anomalous interactions have been reported before in confined
colloidal monoalyers
\cite{kepler94,crocker96a,carbajaltinoco96,Han03b}, but their origin
has remained unresolved for more than a decade.
Is this minimum a real albeit unexplained
feature of the charged colloids' pair interaction,
or is it an artifact of the measurement technique?

\subsection{Temperature-based consistency tests for $u(r)$}
\label{sec:application1}

Because hyperconfigurational temperatures depend sensitively 
on particles' interactions, they constitute a
hierarchy of thermodynamic self-consistency tests for $u(r)$.
The first successful application of the configurational temperature formalism
to colloidal interaction measurements was reported in Ref.~\cite{Han04a}.
Here, we provide a more detailed description of these tests'
implementation and compare their
results with those from thermodynamic sum rules.
Together, this suite of independent tests confirms that pair potentials
extracted from digital video analysis of confined charge-stabilized colloidal
monolayers can be both thermodynamically self-consistent and accurate.

\begin{figure}[t!]
  \centering
  \includegraphics[width=\columnwidth]{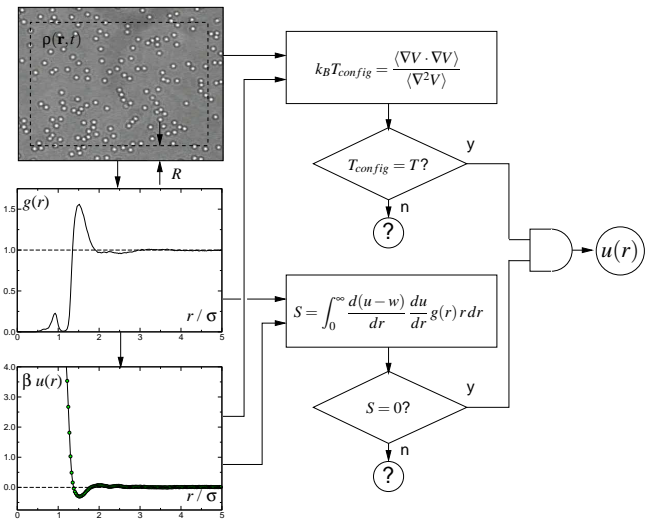}
  \caption{Flow chart for applying configurational temperatures to
    colloidal interaction measurements. 
    From the measured positions of particles, $\rho(\vecr,t)$, we can
    calculate  the radial distribution function $g(r)$ and, from this,
    the candidate pair potential, $u(r)$.  $\rho(\vecr,t)$ and $u(r)$
    are used to calculate the configurational temperatures, while
    $g(r)$ and $u(r)$ are used to compute the sum rule in
    Eq.~(\ref{eq:sumrule}).  Thermodynamic self-consistency in both
    tests suggests that $u(r)$ accurately describes equilibrium
    pair-wise interactions in the system.  Data were obtained for a
    sample at $H = 18~\micron$.}
  \label{fig:flowchart}
\end{figure}

The flow chart in Fig.~\ref{fig:flowchart} outlines our strategy
for checking and optimizing pair potentials.
Given a candidate pair potential, we calculate a variety of
configurational temperatures and compare these with the known
bulk temperature at which the experiment was performed.
Equation~(\ref{eq:oz}) conveniently
yields $u(r)$ in units of $k_B T$, so that configurational temperatures
derived from $u(r)$ are automatically normalized by the bulk thermodynamic
temperature, $T$.
The condition for thermodynamic consistency therefore reduces to
$T_{config} = 1$ in these units.
Successful collapse of
the hierarchy of configurational temperatures to
the thermodynamic temperature suggests that all of the conditions
assumed in the configurational temperatures' derivations have been met
and that $u(r)$ accurately describes pairwise additive contributions to
the free energy of a homogeneous isotropic system in equilibrium.
Conversely, a failure to converge suggests that one or more
of these conditions has not been met.

Identifying departures from equilibrium is particularly important in
experimental systems where subtle fluid flows or environmental
drifts can lead to correlations that might be mistaken for intrinsic
interactions.
Numerical simulations have demonstrated that the configurational temperature
tracks sudden temperature jumps \cite{Butler98} far faster than the pressure
or internal energy, and that its fluctuations also decay more rapidly,
particularly for shorter-ranged potentials \cite{Ennis01}.
These observations suggest that the configurational temperature may
converge to a well-defined value even before the system achieves equilibrium.
This surprising insensitivity has been ascribed \cite{Ennis01} to the
dominant contribution to the configurational temperature of forces among
nearby pairs of particles, which can relax to near-equilibrium configurations
long before a disturbance can propagate through the entire system.
The configurational temperature, therefore, is a better probe of local
equilibrium than global.

\subsection{A sum rule for interaction measurements}
\label{sec:sumrule}

Thermodynamic sum rules provide the necessary tests for
global equilibrium.
A particularly convenient form may be derived from Eq.~(\ref{eq:Tconfig1})
if the pair potential is radially symmetric.
In this case,
\begin{equation}
  \label{eq:avg0}
  k_B T \avg{\nabla_r^2 u(r)} = \avg{\abs{\nabla_r u(r)}^2},
\end{equation}
and we can explicitly calculate the thermodynamic averages of both sides of
this equation:
\begin{align}
  \label{eq:meanddU}
  \avg{\nabla^2_r u(r)} 
  & = 2\pi n N \, \int_0^{\infty} \left(
    \frac{1}{r}\, \frac{du}{dr}+\frac{d^2u}{dr^2} \right) \, rg(r) \, dr 
  \nonumber\\
  & = 2\pi n N \, \left\{ \int_0^{\infty} 
  \frac{du}{dr} \, g(r) \, dr +
  \left. \frac{du}{dr} \, rg(r) \right|_0^{\infty} \right. \nonumber \\
  & \left. \quad - \int_0^{\infty} 
  \frac{du}{dr} \, \frac{d}{dr} \left(rg(r)\right) \, dr \right\}
  \nonumber\\
  & = 2\pi n N \, \int_0^{\infty} 
  \frac{du}{dr} \, \frac{dg(r)}{dr} \, r \, dr,
\end{align}
and
\begin{equation}
  \label{eq:meandudu}
  \avg{\abs{\nabla_r u(r)}^2} 
  = 2\pi n N \, \int_0^{\infty} 
  \left(\frac{du}{dr} \right)^2 \, g(r) r \, dr.
\end{equation}
Combining these results yields the sum rule
\begin{equation}
  \label{eq:sumrule}
  \int_0^\infty \left(
    \frac{du(r)}{dr} - \frac{d \ln{g(r)}}{dr} \right) \,
  \frac{du(r)}{dr} \, g(r) \, r \, dr = 0.
\end{equation}
This sum rule should apply at arbitrary areal densities for any system
whose interactions can be described by a pairwise-additive central potential, $u(r)$.
A similar result was obtained in Ref.~\cite{Baranyai00} for
three-dimensional systems.

Using the radial distribution function to average over pairs of particles
removes any sensitivity to local structural variations, and thus focuses
attention on global properties such as the degree of equilibration.
Consequently, Eq.~(\ref{eq:sumrule}) complements the hierarchy of
configurational temperature consistency checks.

\subsection{Practical considerations}

\subsubsection{The range of interactions}
These thermodynamic tests turn out to be exceedingly sensitive to
imperfections in experimental data, and care is required to apply them
meaningfully.
For example, video microscopy data necessarily is restricted to a limited
field of view, even if the sample itself is substantially larger.
Particles near the edge of the field
of view may have strongly interacting neighbors just out of sight
whose contributions to their net force would be overlooked.
The large apparently unbalanced forces due to 
this pernicious edge effect would
grossly distort estimates of the configurational
temperature were they included in averages such as Eq.~(\ref{eq:TconfigF}).

\begin{figure}[htbp]
  \centering
  \includegraphics[width=.8\columnwidth]{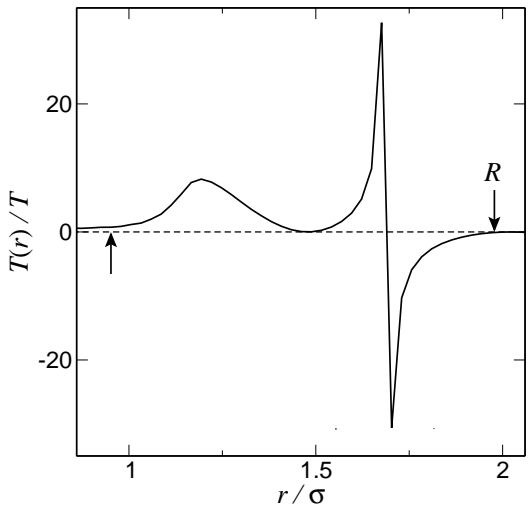}
  \caption{Contributions to the configurational temperature due to pairs of 
    colloidal spheres separated by distance $r$, calculated from the fifth-order
    polynomial fit to the data in
    Fig.~\ref{fig:u(H=9)}.  For this system, spheres separated
    by more than $R = 2 \sigma$ contribute negligibly to the configurational
    temperature.}
  \label{fig:trange}
\end{figure}

To avoid this, we calculate force distributions only for particles whose
relevant neighbors all lie within the field of view. 
Such particles lie no closer than the
interaction's range $R$ to the edge of the field of view. 
We estimate $R$ from $u(r)$ and $g(r)$ by computing
\begin{equation}
  \frac{T(r)}{T} = 
  2\pi \frac{r}{\sigma} g(r) \,
  \frac{\abs{\nabla \beta u(r)}^2}{\nabla^2 \beta u(r)},
\end{equation}
which heuristically describes an effective contribution to the configurational 
temperature of $2 \pi r g(r)$ pairs of spheres interacting with potential $u(r)$
at range $r$ \cite{Han04a}.
A typical example appears in Fig.~\ref{fig:trange}.
This should be considered no more than a heuristic guide because it neglects
three-particle correlations, which are known \cite{Ennis01}
to be important for estimating the temperature.

Pairs at large enough separation that their interactions are vanishingly
weak contribute negligibly to $T(r)$.
The interaction range $R$ therefore can be estimated from the trailing
edge of $T(r)$ in Fig.~\ref{fig:trange}.
By considering particles no closer than $R$ to the edge of the field of view, we
can ensure that no relevant pair interactions will be missed in calculating
the full configurational temperature.
The restricted field of view is indicated by the 
small rectangle overlaid on the
photograph in Fig.~\ref{fig:flowchart}.
Given this cut on the data, we can proceed to calculate the full configurational
temperature.

The data in Fig.~\ref{fig:u(H=9)} display two features whose validity we
can assess using the configurational temperature formalism.  The first
is the anomalous minimum that can be interpreted as evidence for long-ranged
attractions between like-charged particles.  The second is the surprisingly
weak contact repulsion.
This low barrier to aggregation probably is not real, 
otherwise particles would aggregate rapidly by van der Waals attraction. 
Could the minimum similarly be an artifact?

Closer inspection of the data in Fig.~\ref{fig:u(H=9)}
reveals that $g(r)$ is finite even at $r < \sigma$, which should be impossible.
Two sources of experimental error, projection error due to particles' small
out-of-plane motions and polydispersity in the spheres' diameters, 
artificially increase $g(r)$ near contact and thus dramatically reduce 
the apparent interaction energy near $r = \sigma$. 
What Fig.~\ref{fig:u(H=9)} shows is the \emph{effective potential} 
consistent with the raw data for the particles' positions $\rho(\vec{r},t)$, 
including both of these contributions. 
The configurational temperature calculation also is based on the same 
raw position data and so requires
the associated \emph{effective potential} as an input.
The consistency condition $T_{config} = 1$ thus tests the
accuracy and thermodynamic self-consistency of $u(r)$ for the measured
set of $\rho(\vecr,t)$ data.

Given $\rho(\vecr,t)$ and $u(r)$, the net force $\force_j(t)$ on the $j$-th
particle at time $t$ can be estimated using Eq.~(\ref{eq:netforce}), with
the sum over neighboring
particles being restricted to those with $\abs{\vecr_i - \vecr_j} \leq R$.
The set of single-particle forces then can be compiled into estimates for the
configurational temperature for that particular snapshot, and a sequence
of snapshots averaged to obtain a final result.
Averaging is not necessary if each snapshot
captures a large enough number of particles.
The dilute samples in our study, however, typically yield $N = \order{100}$
so that several thousand frames are required for adequate statistics.

\subsubsection{Influence of measurement errors}

\begin{figure}[t!]
  \centering  
  \includegraphics[width=\columnwidth]{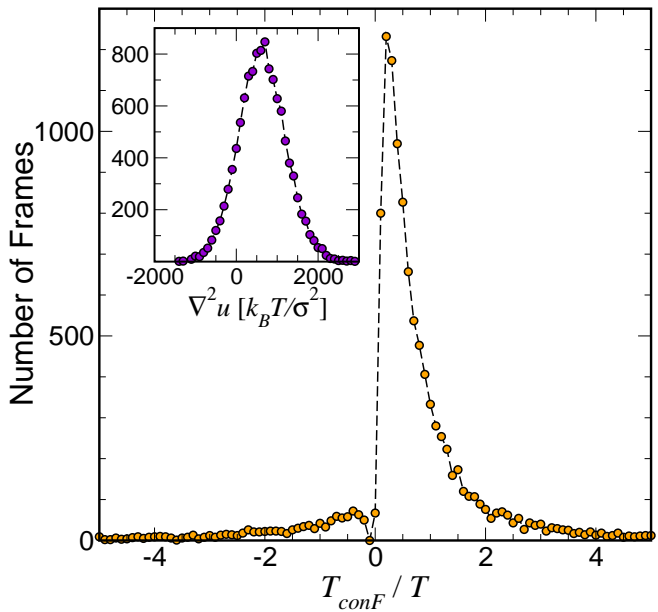}
  \caption{Histogram of the configurational temperatures calculated from the measured
    pair potential in Fig.~\ref{fig:u(H=9)} with the interaction range
    $1< r/\sigma <10$ for a set of 11912 
    video frames, with an average of $N = 288$ particles per frame.
    Inset: corresponding histogram of $\sum_i \nabla^2 u_i$.}
  \label{fig:T9histraw}
\end{figure}

Even when care is taken to avoid edge effects,
naively calculating forces, and thus temperatures, with the
experimentally sampled $u(r)$
yields unsatisfactory results,
as the data in Fig.~\ref{fig:T9histraw} demonstrate.
This shows the histogram of $T_{conF}$ values obtained from the
11912 video frames used to generate Fig.~\ref{fig:u(H=9)}. 
Although the distribution is indeed peaked around unity,
the average, $T_{conF} = 1155$, deviates wildly from the expected value.
Other definitions also fare badly, with $T_{con1} = 23$, for example.
These extraordinarily large averages are not representative
of the force distribution, however.
Rather, they can be ascribed to a small number of frames with huge
apparent temperatures. 
More disturbingly, many
frames appear to have negative temperatures.

\begin{figure}[t!]
\centering 
\includegraphics[width=\columnwidth]{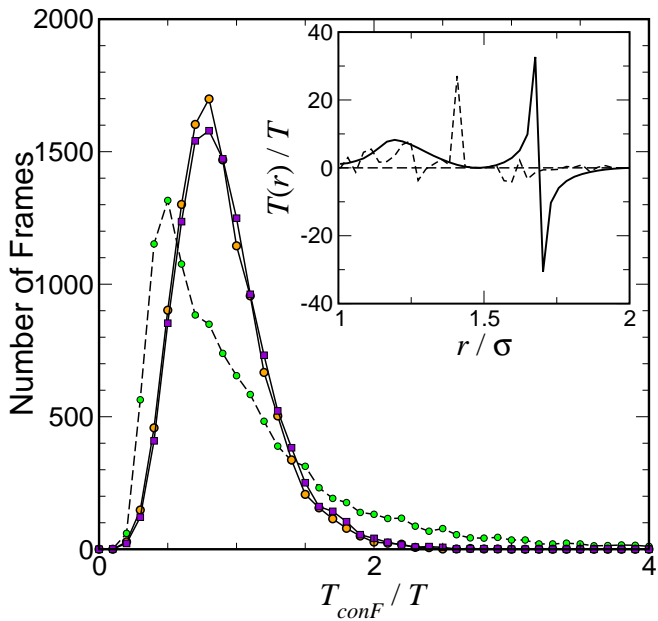}
\caption{Histogram of the configurational temperatures along the $x$ and $y$
  directions calculated with the polynomial fit to the pair potential in
  Fig.~\ref{fig:u(H=9)}.  Solid curves: $T_{conF}^{(x)}$ (circles) and
  $T_{conF}^{(y)}$ (squares).  Dashed curve: $T_{conF}$ from raw $u(r)$.
  Inset: Contribution to the configurational temperature due to
  pairs at separation $r$ for the measured $u(r)$ (dashed curve) and the
  polynomial fit to $u(r)$ (solid curve).}
  \label{fig:Thist}
\end{figure}

Frames with negative temperature
must have $\sum_j \nabla^2 u_j < 0$.
The inset to Fig.~\ref{fig:T9histraw} shows that this is indeed the case.
Such negative values should be extraordinarily rare in an equilibrated
system because they signal mechanical instability.
In this case, however, they can be ascribed to scatter in the
experimentally determined potential, $u(r)$, which is greatly
emphasized by the Laplacian operator.
Particularly for $r > 2 \sigma$ where $u(r)$ is close to 0,
the signal-to-noise ratio becomes small and errors in the computed
configurational temperature becomes unacceptably large. Cutting the
long-range part of $u(r)$ helps to some degree by substantially
decreasing the number of frames with negative temperature. 
However,
this does not suppress the high-temperature tail
in the histogram of single-frame
temperatures, as can be seen from the dashed plot in Fig.~\ref{fig:Thist}. 
These artifacts still boost the apparent temperature to an unrepresentative
$T_{conF} = 1150$. 

To minimize the influence of scatter in $u(r)$, we compute the
configurational temperature using a
fifth-order polynomial fit shown in Fig.~\ref{fig:u(H=9)} 
as the input potential.
This simple smoothing procedure eliminates both negative and 
unreasonably large values of the calculated configurational temperature.
The solid curves plotted in Fig.~\ref{fig:Thist}. 
show the resulting distribution of configurational temperatures factored
along the orthogonal $x$ and $y$ directions in the field of view
according to Eq.~(\ref{eq:tcomponents}).
The two component, $T_{conF}^{(x)}$ and $T_{conF}^{(y)}$ differ by less than
2.5\%, which helps to confirm the system's isotropy.
Their mean, $T_{conF} = 0.969$, also is reassuringly close to the
expected value.

Even when the temperature is calculated with the smoothed pair
potential,
a few frames still have negative temperatures.
This is reasonable if $\lim_{r \to \infty} u(r) = 0$ and the
pair potential has an extremum at an intermediate separation,
$0 < r_0 < R$.
In this case $\nabla^2 u_j$ must change sign and some particles
may contribute negative values to the average in a snapshot.
Chances improve for the average itself to be negative if $N$ is small.
For example, 10 out of 11912 frames have negative apparent temperatures
in a sample with $\avg{N} = 97$.
Negative temperatures are not observed in samples characterized
by purely repulsive interactions, such as the example introduced
in Sec.~\ref{sec:imperfections}.

\subsubsection{Finite size scaling}
The relatively small number of particles in the field of view has
other ramifications.
Because the various configurational temperature definitions involve
different approximations of order $1/N$, we might expect their
results to differ from each other and from the actual thermodynamic
temperature accordingly.
Imaging a substantially larger region of the sample is not feasible.
On the other hand, deliberately sub-sampling the field of view allows us
to probe the dependence on sample size, for which we can extrapolate 
the configurational temperatures to the
thermodynamic limit.

\begin{figure}[t!]
  \centering 
  \includegraphics[width=\columnwidth]{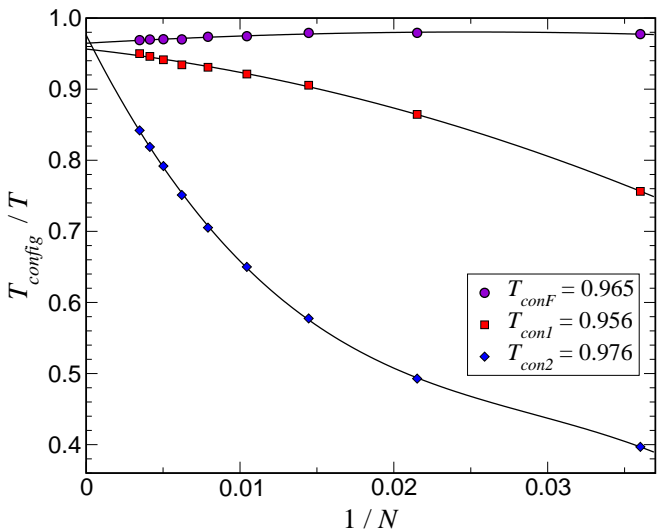} 
  \caption{Finite size scaling of $T_{conF}$, $T_{con1}$ and $T_{con2}$
    for the data in Fig.~\ref{fig:Thist}, with area-weighted fits to 
    second-, second- and third-order
    polynomials, respectively.}
  \label{fig:T9fit}
\end{figure}

The data in Fig.~\ref{fig:T9fit} were obtained by
calculating the configurational temperature with the
$640 \times 480~\unit{pixel}^2$ field of view reduced by borders of 20, 40,
60, 80, 100, 120, 140, 160 and 180 pixels.
The interaction range in this system is 
$R = 2\sigma = 14.9~\unit{pixel}$.
Reducing the number of particles in this manner substantially changes the
estimated temperatures, thereby confirming the importance
of finite-size scaling.
We fit
these results to polynomials in $1/N$ with statistical weighting
estimated from the area and the interaction range.
These weighted fits describe the data very well, so that
the extrapolation to $1/N=0$ should yield meaningful
estimates for the configurational temperatures in the thermodynamic limit. 
Indeed, the
extrapolated results, $T_{conF} = 0.965$, $T_{con1} = 0.956$ and
$T_{con2} = 0.976$, agree quite well with each other,
 and all appear to be
consistent with unity.
Confirming thermodynamic consistency, however, requires us to assess the
configurational temperatures' sensitivity to errors in $u(r)$.

\subsubsection{Sensitivity to input potential}
The apparently good extrapolation of the configurational temperatures 
to the thermodynamic temperature would offer little insight into the
nature of the confined colloids' interactions if these results were insensitive
to relevant features in the pair potential.
For example,
considerable attention has been paid in the literature to the
possibility that anomalous confinement-induced like-charge
attractions such as the example in Fig.~\ref{fig:u(H=9)} might
be artifactual. 
However, if we truncate the negative region of $u(r)$ to create
a purely repulsive potential
and recalculate the configurational temperatures, 
$T_{conF}$, $T_{con1}$ and $T_{con2}$
all extrapolate to 1.5, an error of $150^\circ\unit{C}$.
The large deviation resulting from this admittedly crude test
suggests that the observed attraction is indeed an
integral part of the charged particles' interaction.

\begin{figure}[t!]
  \centering
  \includegraphics[width=\columnwidth]{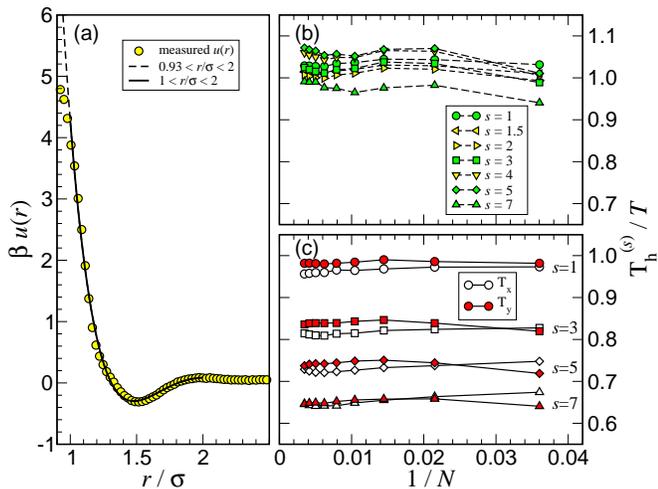} 
  \caption{(a) Measured pair potential $u(r)$ together with
  least-squares fits to fifth-order polynomials over the range 
  $0.93 < r/\sigma < 2$ and $1 < r/\sigma < 2$.
  (b) Hyperconfigurational temperatures including data from $r <
  \sigma$.  $\Th{1} = 1.012$.
  (c) Hyperconfigurational temperatures over the more restricted
  range, factored into Cartesian components.}
\label{fig:hyperT9conF}
\end{figure}

More sensitive tests for thermodynamic consistency are provided by
the hyperconfigurational temperatures defined in Eq.~(\ref{eq:hyperT}).
Figure~\ref{fig:hyperT9conF} demonstrates how small variations in $u(r)$
can cause the hyperconfigurational temperatures to deviate
with respect to each other and also with respect to the thermodynamic
temperature.
Two smoothed versions of the potential are plotted in
Fig.~\ref{fig:hyperT9conF}(a), one fit over the range
$0.93~\sigma \le r \le R$ and the other over the more restricted
$\sigma \le r \le R$.
The former collapses the entire hierarchy of hyperconfigurational
temperatures plotted in Fig.~\ref{fig:hyperT9conF}(b)
to $\avg{\Th{s}}_s \approx 1$ in the extrapolated
thermodynamic limit, with $\Th{1} = 1.012$, which
compares favorably to $T_{conF}=0.965$ in Fig.~\ref{fig:T9fit}.
The latter yields the far less satisfactory results in
Fig.~\ref{fig:hyperT9conF}(c).
Rather than collapsing onto the thermodynamic temperature,
$\Th{s}$ deviates systematically to lower values with larger index
$s$.

This qualitative difference is due to substantial contributions
from pairs of particles with $r < \sigma$.
Such pairs should not be present in a monodisperse sample of impenetrable
spheres, but appear in practice because of the sample's 3\%
polydispersity in radius and because of projection errors due to the
particles' out-of-plane fluctuations.
These two effects are responsible for the observed correlations 
at $r < \sigma$ in Fig.~\ref{fig:u(H=9)}, and for the unreasonably
small values of $u(r)$ in the unphysical range $0.5~\sigma < r < \sigma$.
The successful collapse of the configurational and
hyperconfigurational temperatures under these conditions 
demonstrates that the
\emph{effective} potential accounts
for the apparent particle distribution $\rho(\vecr)$
and may differ subtly from the \emph{ideal} pair potential.

The results in Fig.~\ref{fig:hyperT9conF}(b) and
\ref{fig:hyperT9conF}(c) reflect the general
trend that higher order
hyperconfigurational temperatures are more sensitive to 
details of the input
potential. 
Even so, we can adjust the input potential 
within the experimental error bounds so that all of the
hyperconfigurational temperatures converge to unity. 
In this sense, the hyperconfigurational temperatures not only strengthen our
conclusions regarding the nature of anomalous like-charge attractions, 
but also enable us to improve our estimates for $u(r)$ by adjusting
for improved thermodynamic self-consistency.

The data in Fig.~\ref{fig:hyperT9conF}(c) also highlight another
general feature of the configurational temperatures.  Even though
the more restricted trial potential does not successfully collapse the
data, it does yield consistent results for configurational
temperatures factored along orthogonal directions.  This is a good
indication that, indeed, the system is isotropic.

\subsubsection{Checking for isotropy}

Comparing temperatures measured along orthogonal directions
not only provides a useful check for the interactions' isotropy,
but also can be used to appraise the imaging system.
For an isotropic sample, we expect that 
\begin{equation}
  \frac{\Delta T}{\avg{T}} \equiv 2 \, \frac{T_x - T_y}{T_x + T_y}
\end{equation}
will vanish for \emph{arbitrary} choices of $u(r)$. 
More usually, the combination of a commercial video camera and
video frame grabber results in slightly different length scale
calibrations.
Even quite small differences measurably affect the temperatures calculated
from the associated force components if the interaction is assumed
to be isotropic.
For example, we found that
$\Delta T = -0.025~\avg{T}$ for the confined sample at $H = 9$
and $\Delta T = 0.06~\avg{T}$ for the sample at $H = 195~\micron$.
Even such small differences are plainly visible in 
Fig.~\ref{fig:hyperT9conF}(c). 

Apparent anisotropies of this magnitude
appear consistently in our data sets regardless of
the samples' composition, concentration, degree of confinement,
and so are unlikely to reflect statistical errors.
Nor are they likely to signal a real anisotropy in our
samples' interactions.
Instead, they result from the hyperconfigurational temperatures'
sensitivity to subtle geometric distortion in our imaging system.
Rescaling the measured $x$ and $y$ coordinates
slightly can substantially reduce the apparent anisotropy
in the entire hierarchy of hyperconfigurational temperatures, 
as the data in Table~\ref{tab:anisotropy} show.
For the system used in this study, a 0.7\% correction of the $x:y$ scale ratio is 
enough to account for the 5.8\% anisotropy of 
$\Th{1}$ in the $H = 195~\micron$ data.

The same scaling factor also corrects the apparent anisotropy
in the other samples we have studied, and thus appears to be
correctly interpreted as a correction to the calibration of our
imaging system.  
Furthermore, differences in the scaling factors as small as
$\pm 0.1\%$ perform substantially less well, as shown by the data 
in Table~\ref{tab:anisotropy}.
This level of sensitivity greatly exceeds the typical 1\%
calibration accuracy obtained by imaging test patterns,
and thus provides a new tool for assessing and correcting
geometric defects in the digital video microscopy system.

\begin{table}
  \centering
  \begin{tabular}{|c|c||c|c|c|c|}
    \hline 
    \multicolumn{2}{|c||}{} & \multicolumn{4}{c|}{\textbf{$x:y$ scaling factor}}\\
    \cline{3-6}\multicolumn{2}{|c||}{$\frac{\Delta T}{\avg{T}}$}
    & 1:1 & 1.003:0.997 & 1.003:0.9963 & 1.003:0.996 \\ \hline \hline 
    & 1 & 5.8\% & 0.61\% & -0.07\% & -0.82\% \\ 
    \cline{2-6}$s$ & 3 & 4.4\% & 0.87\% & 0.009\% & -0.20\% \\
    \cline{2-6} & 5 & 2.6\% & 1.3\% & 0.16\% & 0.90\% \\ 
    \cline{2-6} & 7 & 1.4\% & 1.8\% & 0.65\% & 0.71\%\\ \hline
  \end{tabular}
  \caption{Correcting apparent anisotropy in the hyperconfigurational
    temperatures of the $H = 195~\micron$ data set
    by rescaling coordinates.  In each case, $g(r)$, $u(r)$
    and $\Th{s}$ were recalculated with revised particle locations.}
  \label{tab:anisotropy}
\end{table}

Successfully correcting apparent anisotropy in the hyperconfigurational
temperatures also provides insight into the nature of the system's
interactions.
If replacing $u(r)$ by an arbitrary function causes 
$T_x$ to differ from $T_y$ then the system indeed may be anisotropic,
either because its interactions are anisotropic, or
else because of its response to an external field.
In the latter case, the external field contributes an additional
configuration-dependent term to the Hamiltonian, $\ham_{ext}(\vecG)$,
which contributes, in turn, to the definitions of the configurational
temperatures.  
If $\partial_x \ham_{ext}(\vecG) \neq \partial_y \ham_{ext}(\vecG)$,
then the configurational temperatures along the two directions
generally will differ.

\subsubsection{Accounting for sample imperfections in $\Th{s}$}
\label{sec:imperfections}

\begin{figure}[t!] 
  \centering
  \includegraphics[width=\columnwidth]{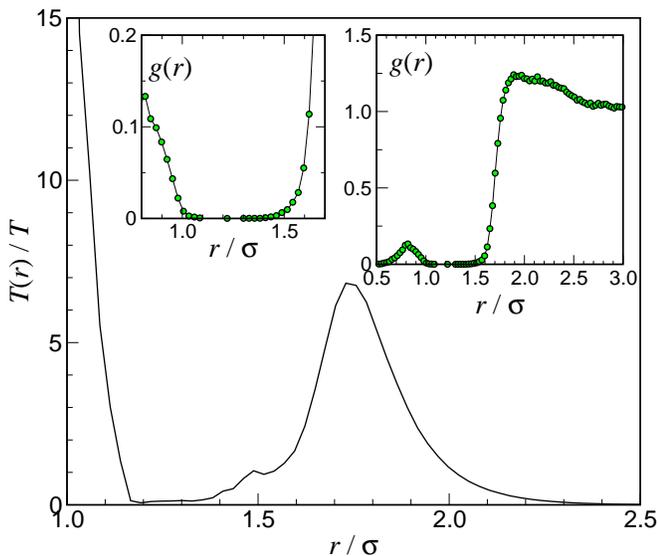}
  \caption{Contribution to the configurational temperature at different pair 
    separations assuming interactions described by Eq.~(\ref{eq:DLVO})
    with $\kappa^{-1} = 180~\unit{nm}$, $Z = 7563$ for a silica
    monolayer at $H=195~\micron$. 
    The peak at $1 < r/\sigma < 1.2$ is caused by dimers. 
    The tiny peak at $r = 1.5~\sigma$ is due to one particle stuck to the
    bottom surface.
    Right Inset: $g(r)$ with a dimer peak around $r = 0.8~\sigma$.
    Left Inset: a more detailed view of $g(r)$ near $r = \sigma$.}
  \label{fig:rgT}
\end{figure}

Colloidal samples quite often include small populations of
aggregated pairs of spheres, also known as dimers.
These can have a striking effect on the configurational
temperature, as the data in Fig.~\ref{fig:rgT} demonstrate.
This sample also consists of $\sigma = 1.58~\micron$ diameter
silica spheres, but
sedimented to the bottom wall of
a slit pore $H = 195~\micron$
thick at an areal density $n\sigma^2 = 0.080$.
Unlike samples in relatively thin ($H = 9~\micron$) slit pores
considered so far, such weakly confined silica monolayers are
found to have purely repulsive screened-Coulomb interactions \cite{Behrens01a,Han03b}
in at least qualitative agreement with mean-field 
theory.
The presence of a few dimers, however,
contribute a small peak to the radial distribution
function at $r = 0.8~\sigma$ whose tail extends to $1.1~\sigma$.
These excess correlations are dramatically magnified
in $T(r)/T$ because any two 
nominally repulsive particles 
ought to exert exceptionally large forces on each other near
contact.
The overall result is extremely large configurational temperatures.

At first glance, such defects in the sample would appear to
render meaningful measurements of the configurational temperature impossible.
However, the function
$T(r)/T$ can be used to cut the spurious data while retaining
enough useful information for an accurate assessment.
In particular, the peak in $T(r)/T$ at small $r$ is well separated
from the principal peak at $r = 1.8~\sigma$.
This suggests that the former can be ascribed entirely to dimers
and the latter to genuine long-ranged interactions, with a clean
division at about $r = 1.3~\sigma$.
It seems reasonable, therefore, to eliminate dimers' contribution
to the configurational temperatures by excluding any pair separations
smaller than $1.3~\sigma$.

Restricting the range of $u(r)$ too much would exclude relevant interactions,
causing the net force on many particles to appear unbalanced
and the configurational temperature to increase accordingly.
For the present data set, we find that increasing the lower cutoff
to $r = 1.45~\sigma$ only increases the apparent temperature by a few
percent. This indicates that three-body correlations are weak at this
concentration range and that the proposed cutoff at $r = 1.3~\sigma$ will
not distort the results. Similarly, ignoring contributions from pairs
separated by more
than $r = 2~\sigma$ has little influence on the estimated
temperature and establishes the interaction range for accurate
temperature estimates.

\begin{figure}[t!]  
  \centering
  \includegraphics[width=\columnwidth]{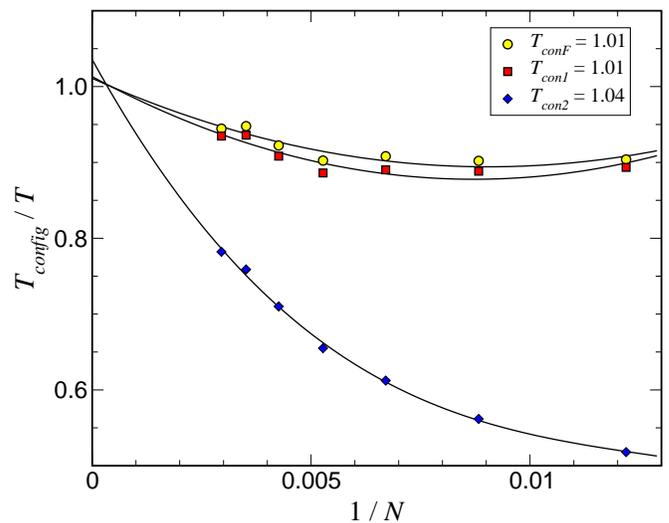}  
  \caption{Finite size scaling of three configurational temperature definitions
    computed for a purely repulsive monolayer of 
    $\sigma = 1.58~\micron$ silica spheres in a
    slit pore of height $H=195~\micron$. 
    Solid curves are weighted fits to second-order polynomials in $1/N$ used
    to extrapolate to the thermodynamic limit.
    Data at $1/N < 0.005$ are influenced by the stuck particle, as
    discussed in the text.}
  \label{fig:TsizeH195}
\end{figure}

\begin{figure}[t!]
  \centering
  \includegraphics[width=\columnwidth]{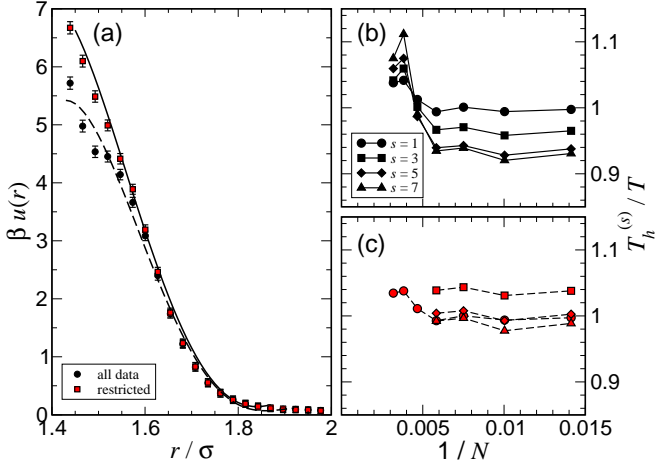}
  \caption{Influence of an immobilized sphere on the effective pair potential and
    hyperconfigurational temperatures.  (a) Measured pair potentials together with
    fourth-order polynomial fits yielding optimal collapse of the hyperconfigurational
    temperatures.  Squares: results including all data.  Circles: results obtained
    by excluding the region around the stuck sphere.
    (b) Hyperconfigurational temperatures obtained for the entire field of view.
    (c) Hyperconfigurational temperatures obtained for the restricted data 
    set excluding the region around the stuck particle.}
  \label{fig:hyperT195conF}
\end{figure}

Figure~\ref{fig:TsizeH195} shows typical finite-size scaling results for
a trial fourth-order polynomial fit to the experimentally obtained
pair potential over the range $1.435 < r/\sigma < 1.89$.
The fit potential is plotted as a dashed curve in
Fig.~\ref{fig:hyperT195conF}(a).
Replacing this with a fit to the predicted mean field potential
described by Eq.~(\ref{eq:DLVO}) over the range 
$1.2 < r/\sigma < 1.95$
yields comparably good convergence, and increasing the range
to $1.2 < r /\sigma < 10$ decreases the extrapolated configurational
temperature by just 4 \%.
All of these trial potentials fall within error estimates for
the measured potential.
On this basis, and because the various definitions of the
configurational
temperature all extrapolate to unity,
we conclude
that the measured potential once again accurately describes the
system's equilibrium pair interactions.
In this case, however, the potential appears to be purely repulsive.

The sample used to compile these data also included one particle that had
deposited irreversibly onto the lower glass surface near the edge of the field of view.
A single immobilized sphere might not be expected to influence the free monolayer's
structure and dynamics much.
The influence on the radial distribution function is indeed subtle,
with the slight peak at $r = 1.5~\sigma$ in $T(r)/T$ (Fig.~\ref{fig:rgT})
disappearing when the region containing the stuck particle is excluded
from the calculation.
The effect on the candidate pair potential is somewhat more pronounced, particularly
for $r < 1.5~\sigma$, as can be seen in
Fig.~\ref{fig:hyperT195conF}(a).
But does the potential from the restricted data set better reflect the
system's interactions?
After all, the configurational temperatures calculated with the
unrestricted data in Fig.~\ref{fig:TsizeH195} all extrapolate
reasonably well to the thermodynamic value.
Figures~\ref{fig:hyperT195conF}(b) and \ref{fig:hyperT195conF}(c) 
show that the hierarchy of hyperconfigurational
temperatures collapses to unity only when the region
containing the stuck particle is excluded.
The restricted data set, therefore, offers a more accurate picture
of the pair potential.
This is consistent
with our earlier observation that small uncertainties in $g(r)$ at
small separations can be dramatically magnified in
$u(r)$.

\subsection{Applying the thermodynamic sum rule}

\begin{figure}[t!]
  \centering
  \includegraphics[width=\columnwidth]{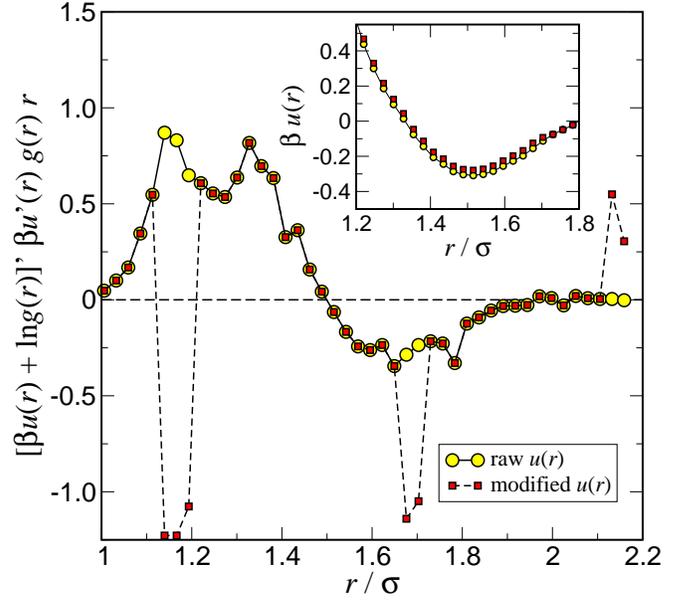}  
  \caption{Integrand of the sum rule in Eq.~(\ref{eq:sumrule}) for the confined
    silica monolayer at $H=9~\micron$.
    Inset: Best fit pair potential from Fig.~\ref{fig:u(H=9)} (circles) and another
    estimate (squares) consistent with both the confidence interval of $u(r)$ and with
    convergence of the configurational temperatures to the thermodynamic value.
    The former gives a relative error of 0.5 in the sum rule, and the latter 0.001.
    }
  \label{fig:sumruledemo}
\end{figure}

Even when the
configurational and hyperconfigurational temperatures
converge to the thermodynamic value for a reasonable choice of pair potential, 
the system's degree of equilibration still has to be assessed.
Figure~\ref{fig:sumruledemo} shows the integrand of 
the sum rule in Eq.~(\ref{eq:sumrule})
for two slightly different effective potentials, both of which are
consistent with the interactions measured in
the confined colloidal monolayer at $H = 9~\micron$.
Even differences in the input potential too small to affect
the configurational temperatures
can change the sum rule's integrand substantially.
These changes affect whether or not the sum rule as a whole
is satisfied.

Normalizing the integral in Eq.~(\ref{eq:sumrule}) by the integral of the
absolute value provides a useful measure of convergence.
The best-fit pair potential obtained from the liquid structure
inversion (plotted as circles in the inset to
Fig.~\ref{fig:sumruledemo})
yields an unacceptably
large relative error of 0.5.
At such low areal densities ($n \sigma^2 = 0.0684$), $u(r)$
is very similar to the potential of mean force, $w(r) = - k_B T \, \ln g(r)$, and
very small changes in $u(r)$ can influence the sum rule's 
integrand substantially.
In fact, adjusting the potential just slightly to the function 
plotted as squares in Fig.~\ref{fig:sumruledemo}
improves the sum rule's convergence to 0.001.
This modified potential still successfully collapses the
configurational and hyperconfigurational temperatures, and thus may
be considered an improved estimate for the pair potential.
This successful application of the thermodynamic sum rule
suggests that the system is indeed in equilibrium, and that the candidate
pair potential, including its long-ranged attraction,
accurately describes the confined particles' interactions.

\subsection{Areal pressures}

The foregoing considerations serve to confirm that the same colloidal
silica spheres experience qualitatively different equilibrium pair interactions
when confined between parallel glass walls separated by $H = 195~\micron$
and $H = 9~\micron$, with the more strongly confined dispersion exhibiting
anomalous long-ranged attractions.
Such a qualitative difference in the system's pair potential also should
manifest itself in the system's other thermodynamic properties, such as its
pressure.
\begin{figure}[t!]
  \centering 
  \includegraphics[width=\columnwidth]{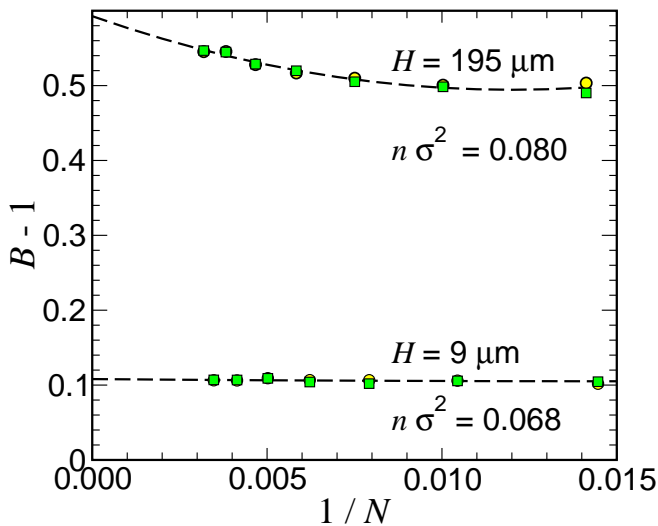}
  \caption{Finite-size scaling of the normalized areal pressure
    for systems with
    repulsive and attractive potentials.}
  \label{fig:VirialP}
\end{figure}

Given the particles' locations and an estimate for their pair interactions,
we can calculate the monolayers' pressure as
$P = n k_B T + p^{pot}$, where $p^{pot}$ is a departure from ideal
gas behavior due to the particles' interactions.
This so-called ``potential pressure'' is give by
\begin{equation}
  \label{eq:pressure} 
  Vp^{pot} = \frac{1}{d} \, \avg{\sum_i \vecr_i \cdot \force_i}
  = \frac{1}{d} \, \sum_{i<j} \vecr_{ij} \cdot \force_{ij}
\end{equation}
where $d$ is the dimension of the space, $\vecr_{ij} = \vecr_j - \vecr_i$
and $\force_{ij} = - \nabla_i u(\vecr_{ij})$. 
The former definition based on absolute coordinates works best for
systems with open boundaries, which are described in the grand canonical ensemble,
but fails in systems with periodic boundary conditions, 
which often are used in computer simulations. 
The latter definition based on relative coordinate works in both.
Either should apply to our experimental data.

Comparing the different systems' interaction-driven departures from
ideal gas behavior is facilitated by defining
\begin{equation}
  \label{eq:virial}
  B = \frac{\beta P}{n} = 1 + \frac{\beta p^{pot}}{n},
\end{equation}
whose dependence on system size is plotted in Fig.~\ref{fig:VirialP}.
These data were obtained with the definition based on absolute
coordinates measured from the center of the field of view.
Equivalent results obtained with relative coordinates differ by less than
1.5\%.
As for the configurational temperature definitions, Eq.~(\ref{eq:pressure})
applies in the thermodynamic limit, and $B$ is obtained by
extrapolating to large system size.
In this case, we see that the system at $H = 195~\micron$ with
long-ranged repulsive
interactions has a substantially higher pressure than would be expected for
an ideal gas at the same areal density.
The increase is much smaller in the more strongly confined system,
presumably because of the
potential's long-ranged attractive
tail.

Like the configurational temperature,
the pressure also depends sensitively on the input potential and
length scale calibration.
Without appropriate rescaling, $P_x$ is 10\% higher than
$P_y$ in the $H = 195~\micron$ data set. 
After rescaling using the factors obtained by requiring
isotropy in the configurational temperatures, 
the two components agree well, as can be seen in
Fig.~\ref{fig:VirialP}. 
This further confirms that the observed
anisotropy is due to artificial image distortion that can be corrected
with a single rescaling factor.

\subsection{Using $\Th{s}$ to measure $u(r)$}
\label{sec:application2}

\begin{table*}
  \centering
  \begin{tabular}{|c|r|c|c|c|c|}
    \hline
    $\kappa^{-1}$ (nm) & $Z$ & $\Th{1}$ & $\Th{3}$ & $\Th{5}$ & $\Th{7}$\\\hline\hline
    160 & 10147 & $1.00 \pm 0.03$ & $1.62 \pm 0.12$ & $1.53 \pm 0.09$ & $1.23 \pm 0.06$ \\\hline
    180 &  6755 & $1.00 \pm 0.02$ & $1.31 \pm 0.09$ & $1.31 \pm 0.07$ & $1.08 \pm 0.05$ \\\hline
    200 &  4779 & $1.00 \pm 0.01$ & $1.07 \pm 0.07$ & $1.12 \pm 0.06$ & $0.94 \pm 0.04$ \\\hline
    \bf 210 & \bf 4096 & $\mathbf{1.00 \pm 0.01}$ & $\mathbf{0.98 \pm
    0.06}$ & $\mathbf{1.04 \pm 0.06}$ & $\mathbf{0.88 \pm 0.04}$ \\\hline
    220 &  3549 & $1.00 \pm 0.01$ & $0.90 \pm 0.05$ & $0.96 \pm 0.05$ & $0.82 \pm 0.04$ \\\hline
    250 &  2438 & $1.00 \pm 0.01$ & $0.72 \pm 0.03$ & $0.77 \pm 0.04$ & $0.68 \pm 0.03$ \\\hline
  \end{tabular}
  \caption{Determining the free parameters in a trial potential using $\Th{s}$. 
    Hyperconfigurational temperatures calculated from the screened-Coulomb potential for
    the $H = 195~\micron$ data set over the range $1.3 < r/\sigma <10$. }
  \label{tab:hyperT}
\end{table*}

The configurational and hyperconfigurational temperatures 
show great promise as thermodynamic self-consistency tests
for interaction measurements in soft-matter system.
They also can be used to measure interactions directly, with
the hierarchy of definitions providing a set of simultaneous
constraints on free parameters in a model for the pair potential.
Even when the theoretical form of the potential is
not known \emph{a priori}, a numerically determined function
that simultaneously collapses the entire hierarchy to the thermodynamic
temperature would provide a model-free estimate for the potential.

As an example of this procedure, we use hyperconfigurational
temperatures to determine the parameters $\kappa^{-1}$ and $Z$
in the screened-Coulomb potential,
Eq.~(\ref{eq:DLVO}), that accounts for the distribution of particles
in the $H = 195~\micron$ data set \cite{Han03b}.
Requiring $\Th{1} = \Th{3} = 1$ yields a screening length
of $\kappa^{-1} = 208.4~\unit{nm}$ and an effective charge on
the particles of $Z = 4123$.
These values also converge $\Th{5}$ and $\Th{7}$ close to unity,
suggesting that the screened-Coulomb potential reasonably
describes the interactions in this system.
Table~\ref{tab:hyperT} shows typical results for parameters
near the optimal values.

Simply fitting Eq.~(\ref{eq:DLVO})
to the measured pair potential 
in Fig.~\ref{fig:hyperT195conF}(a)
yields $\kappa^{-1} = 180 \pm 10~\unit{nm}$
and $Z = 6500 \pm 1000$.
Although these values differ somewhat from those obtained by
collapsing the configurational temperatures, both are broadly consistent
with the charge regulation theory for
interacting silica surfaces~\cite{Behrens01b}.
Their disagreement and the slight departure from unity of the
highest-order hyperconfigurational temperatures suggests a
small deviation of the true pair potential from the predicted
screened-Coulomb form.
This is reasonable because the nearby bottom wall probably 
affects the interaction somewhat.
No convergence is possible at all for the data at $H = 9~\micron$ because
the pair interactions are not well described by the purely repulsive
screened-Coulomb form.

Using hyperconfigurational temperatures to provide constraints on models for the
pair potential offers advantages over other interaction measurement techniques.
Principally, this method eliminates the intermediate steps of
first measuring $g(r)$ and then inverting it to obtain $u(r)$.
This first step generally requires amassing large amounts of data to obtain accurate
information on correlations at small separations.
The second involves uncontrolled approximations and limits the range of concentrations
over which interactions can be measured.
By contrast, the present approach
requires substantially less data and involves fewer approximations.
Consequently, this approach should be useful for exploring denser, 
more strongly interacting systems.
Its ability to distinguish pairwise and many-body interactions also
should be helpful for such studies.

On the other hand, calculating configurational temperatures involves sums over all pairs
of particles, and must be repeated for sub-samples of varying size to account for finite-size
scaling.
This can be computationally expensive, particularly in model-free
searches.
Implementing fast search algorithms on restricted data sets with additional data being
added only in the final stages of polishing should alleviate this problem.

Failure of such a search to converge may signal a failure of pairwise additivity, a departure from
equilibrium, or else an inappropriate model for the pair potential.  Distinguishing
these would probably require resorting to 
one of the alternate measurement methods.

\section{Conclusion and discussion}
\label{conclusion}

The recently introduced notion of configurational temperature provides
a valuable new tool for assessing experimental systems' thermodynamic
state from static snapshots.
We have introduced a hierarchy of
hyperconfigurational temperatures that emerge
naturally from the generalized
definition of temperature, and have shown that both these 
and the generalized definition of the temperature can be derived
from the classical hypervirial theorem.
Colloidal monolayers provide an ideal experimental test bed
for these new concepts in statistical mechanics.

We have tested the configurational and hyperconfigurational
temperature to within 1\% accuracy using particles' distribution and
pair potentials measured in Ref.~\cite{Han03b}. The effect of finite
system size is clearly observed and accounted for in these
measurements. 
Since the configurational temperatures' derivation
requires pairwise additivity and their computation
depends sensitively on the input potential, they can be used as
thermodynamic self-consistency checks for measured pair potentials.
The configurational temperatures calculated for our experimental
data on confined colloidal silica monolayers 
confirm that our measured potentials are
accurate and that they are consistent with the assumption of pairwise
additivity.

We have found that higher-order 
hyperconfigurational temperatures are increasingly sensitive to errors
in the potential $u(r)$
because of their dependence on higher moments of the
force distribution.
Even so, we have found that we can adjust the pair potentials
within the measured uncertainties to converge the entire hierarchy of
hyperconfigurational temperatures to unity for all of our
data sets. This provides substantial independent evidence that our
directly measured potentials reliably reflect equilibrium pair
potentials for our systems.

A sum rule introduced in Sec.~\ref{sec:sumrule} complements the
information provided by the configurational and hyperconfigurational
temperatures by providing additional insights into the system's degree
of equilibration.
Very small errors in the pair potential, moreover, are dramatically
emphasized by the derivatives in the sum rule's definition.
We find, nevertheless, that the sum rule can be satisfied for the colloidal
samples in our study by adjusting the potential within the range of their uncertainties.

Applying these analytical tools to our system of sedimented colloidal silica spheres
allows us to draw new conclusions regarding the nature of electrostatic interactions
in this deceptively simple system.
Rather than being purely repulsive, as mean-field theory predicts, confined colloids'
interactions can be characterized by a strong and long-ranged attraction.
This result echos those reported more than a decade ago in the first generation
of colloidal interaction measurements \cite{kepler94,crocker94,vondermassen94,crocker96,carbajaltinoco96}.
Now, however, we can assert with confidence that the observed anomalous attractions
constitute pairwise-additive contributions to the systems' equilibrium free energies,
and not from any of the myriad of possible experimental artifacts that have been proposed.
This interpretation is further bolstered by measurements of the areal pressure in these
monolayers, which show clear signatures of their differing interactions.
Resolving the anomaly of confinement-induced like-charge attractions
therefore requires a fresh assessment of the nature of
colloidal electrostatic interactions in simple electrolytes.

Hyperconfigurational temperatures also can be used as a set of
constraints to determine the free parameters in a model for a system's
pair interactions.
We have demonstrated this by determining the two free 
parameters in a screened-Coulomb model for charge-stabilized colloids'
interactions in the repulsive regime.
In principle, the unbounded hierarchy of hyperconfigurational 
temperatures can be used in this way to determine any
pairwise additive potential given the position, particularly 
if that potential can be modeled as a polynomial or comparably simple
function.

One advantage of this method is that it circumvents the uncontrolled
approximations that have dogged other approaches to measuring
macroionic interactions in equilibrium. This method in more general in
that it can be applied at arbitrary particle densities.
Unlike measurement techniques based on the radial distribution
function, $g(r)$, furthermore, configurational temperature measurements
can be applied to inhomogeneous systems.
This new method therefore should be useful in phase separated systems
at equilibrium.

We are grateful to Owen Jepps, Sven Behrens and Brian Koss for helpful discussions.
This work was supported by the Donors of the Petroleum Research Fund of the
American Chemical Society.

%

\end{document}